\documentclass[preprint2]{aastex}

\shorttitle{Identifying Breaks and Curvature in the \texit{Fermi} Spectra of Bright Flat Spectrum Radio Quasars}
\shortauthors{Harris et al.}

\begin{document}

\title{Identifying Breaks and Curvature in the \textit{Fermi} Spectra of Bright Flat Spectrum Radio Quasars}

\author{J. Harris, M. K. Daniel \& P. M. Chadwick}
\affil{Dept. of Physics, Durham University, South Road, Durham, DH1 3LE, UK}
\email{j.d.harris@durham.ac.uk}

\begin{abstract}
Knowing the site of $\gamma $-ray emission in AGN jets will do much for our understanding of the physics of the source. In particular, if the emission region is close to the black hole then absorption of $\gamma $-rays with photons from the broad-line region could become significant. Such absorption is predicted to produce two specific spectral breaks in the $\gamma $-ray spectra of Flat Spectrum Radio Quasars (FSRQs). We test this hypothesis using 3 years of \textit{Fermi} observations of nine bright FSRQs.  A simple power law fit to the spectrum of each source can be significantly improved by introducing a break, but the break energies are inconsistent with those predicted by the double-absorber model.  In some cases the fit can be further improved by a log-parabola.  In addition, by dividing the data from each source into two equal epochs we find that the best description of an object's spectrum often varies between a log-parabola and a broken power law.

\end{abstract}


\section{INTRODUCTION}
The commonly accepted paradigm is that active galactic nuclei (AGN) consist of 
a central black hole surrounded by a torus of dust.  There are also two regions of material called the broad-line region (BLR), which extends to a distance of $<0.1\rm~pc$ from the central black hole for bright AGN \citep{kaspi1996}, and the narrow-line region (NLR), which is located $\thicksim10$~-~$1000\rm~pc$ from the central black hole \citep{alexander2012}.  Some AGN also have a relativistic jet perpendicular to the disk.  If this jet has a small line of sight to the observer the object is termed a blazar.
 
Blazars emit radiation across the whole electromagnetic spectrum including the
$\gamma $-ray regime; for example, the  \textit{Fermi Gamma-ray Space Telescope}
has detected hundreds of blazars at photon energies between $100\rm~MeV$ and $100\rm~GeV$ \citep{abdo2010b}.  The shape of a typical blazar's spectral energy distribution is characterized by two peaks when plotted in units of $\rm log(\nu)$ against $\rm log(\nu F_{\nu})$ where $\nu$ is the frequency of the radiation and $F_\nu$ is the flux at frequency $\nu$. Blazars can be divided into two subclasses, Flat Spectrum Radio Quasars (FSRQs) and BL Lacertae objects (BL Lacs).  BL Lacs tend to have their second peak at higher energies than FSRQs which tend to have a second peak below $\thicksim100\rm~MeV$ \citep{ghisellini2009}. In BL Lacs emission lines from the BLR tend to be weak or not-observed \citep{georganopoulos1998}.

The $\gamma $-ray emission from blazars is assumed to originate in the jet, and observed variability timescales suggest that the $\gamma $-ray emission region is small in size ($\lesssim10^{-3}\rm~pc$) \citep{sbarrato2011}. However, both the region and the mechanism of $\gamma $-ray emission is currently unidentified.  One popular theory is that electrons in the jet emit seed photons via synchrotron emission and inverse Compton scatter these photons up to $\gamma $-ray energies \citep[the synchrotron self-Compton model, see e.g.][]{maraschi1992}.  There is also widely thought to be a contribution from photons that originate outside the jet that become upscattered \citep[the external-Compton model, see e.g.][]{dermer1993}. This contribution is thought to be more important in FSRQs than BL Lacs \citep{ghisellini1998} due to the availability of photons from the BLR for upscattering.  In these models the synchrotron emission accounts for the first, low energy peak in the SED while the Compton emission accounts for the second, high energy peak in the SED.  Alternatively, hadronic components in the jet may be accelerated to high enough energies to emit synchrotron radiation in the $\gamma $-ray regime which accounts for the high energy peak in the SED \citep[hadronic models, see e.g.][]{mucke2003}.  This is still an active area of research \citep[see e.g.][for a short review]{bottcher2010} since it is key to extending our knowledge of the underlying jet physics. In order to construct and test these models it is crucial to understand the $\gamma $-ray emission region and the shape of the $\gamma $-ray energy spectrum being produced.

\citet{abdo2010a} concluded that the spectra of FSRQs differ from simple power law descriptions in the \textit{Fermi} energy range ($\thicksim100\rm~MeV$~-~$300\rm~GeV$).  In particular, they concluded that for the brightest FSRQ source, 3C~454.3, a broken power law gives the most acceptable fit compared with simple power law and curved (log-parabola) forms. \citet[hereafter PS10]{poutanen2010} note that recombination of electrons with nuclei can produce sharp increases in photon density above a specific energy. Therefore the opacity to $\gamma $-rays due to photon-photon ($\gamma$-$\gamma$) pair production would also see a sharp increase above a corresponding energy. PS10 used a simple photoionization model of the BLR clouds and an emission spectrum in the energy range $1\rm~eV$~-~$100\rm~keV$ from \citet{laor1997} to model the opacity the BLR presents to a $\gamma $-ray of given energy.  They concluded that there would be two increases in opacity, one occurring in the region $4\rm~GeV$~-~$7\rm~GeV$ due to He~II recombination and one in the region $19.2\rm~GeV$~-~$30\rm~GeV$ due to H~I recombination.  A further simplification in PS10 predicted that these changes should appear at $4.8\rm~GeV$ and  $19.2\rm~GeV$ respectively.  It was noted that if a simple power law of $\gamma $-rays is subjected to such an increase in opacity then the effect is a change in the spectral index akin to a broken power law. Furthermore, it is suggested that this is the cause of the broken power law spectra found in \citet{abdo2010a} and that a second break should occur at a higher energy. This scenario is termed the `double-absorber model'. In PS10, the authors favor the double-absorber model over fitting a power law with a single break in which the break energy is left as a free parameter. This is a very important result if correct, since absorption of $\gamma $-rays by BLR photons strongly implies a $\gamma $-ray production region that lies within the BLR.

We perform analysis of the \textit{Fermi} data from several bright FSRQs in order to test three spectral forms to determine which gives the best description of the data.  The first is a simple power law

\begin{equation}\label{simplepl}
\frac{dN}{dE}=k\left(\frac{E}{E_0}\right)^{-\Gamma},
\end{equation}\newline where ${dN}$/${dE}$ is the differential photon flux as a function of photon energy, $k$ is a normalization constant, $E_0$ is a normalization energy, which we fixed at $0.5\rm~GeV$, and $\Gamma$ is the spectral index. The second form tested is a broken power law, which describes a sudden change in the spectral index of a power law from $\Gamma_1$ to $\Gamma_2$ at a break energy $E_b$

\begin{equation}\label{brokenpl}
{\frac{dN}{dE}}= \left\{
\begin{array}{lll}
 k\left(\frac{E}{E_b}\right)^{-\Gamma_1}	&	{\rm if}	& E<E_b\\
 k\left(\frac{E}{E_b}\right)^{-\Gamma_2}&	{\rm if}	& E\geq E_b
\end{array}
\right\}.	
\end{equation} \newline%
If notable deviation from a simple power law was found to be present, we tested to see if the break energies were in agreement with those predicted from the double-absorber model and if they were stable over time. The final form tested is a log-parabola, a smoothly varying function that naturally also describes a peak frequency  
\begin{equation}\label{logparab}
\frac{dN}{dE}=k\left(\frac{E}{E_0}\right)^{-\Gamma -\beta\log(\frac{E}{E_0})},
\end{equation}\newline where we again fix the normalization energy to $0.5\rm~GeV$ and $\beta$ is a parameter that defines the curvature of the spectrum. A power law is physically untenable over a sufficiently wide range of energies whereas the log-parabola can naturally describe the curvature expected on either side of the peak in the SED.

We used a larger dataset than was available for the PS10 analysis and also made several improvements to the methods used in PS10. As the authors in that paper noted, PS10 used a simplification in the \textit{Fermi}-LAT's containment angle and effective area which degraded the accuracy of photon selection. Secondly, PS10 used only a single region of space to characterize the energy spectrum of the $\gamma $-ray background for each object rather than a more thorough model of diffuse $\gamma $-ray emission and emission from known nearby point sources. Thirdly, PS10 used a binned analysis which introduced extra degrees of freedom in the placement of the bin edges. These simplifications had a noticeable impact on the precision of the fits.  In the broken power law fits performed in PS10 the uncertainties in the break energy were on average $50\%$ larger than those for the same objects over the same time period performed in \citet{abdo2010a}.

Our improvements to this analysis are detailed in Section~2 along with the selection cuts we applied to the data, how we tested for breaks and curvature in the spectrum, and whether any breaks were stable over time.  Our results are presented in Section~3 and our conclusions are presented in Section~4.

\section{Data Analysis}
\subsection{Event Selection}
We analyzed the $9$ FSRQs which were selected in PS10 for their brightness and lack of contamination from other point sources or the Galactic bulge.  These sources are listed in Table~1.

We examined all photons that arrived within a $15^{\circ}$ radius of the target using the P6\_V3\_DIFFUSE version of \textit{Fermi}'s response function. Following standard practice\footnote{
http://fermi.gsfc.nasa.gov/ssc/data/analysis/scitools/likelihood\_tutorial.html
}, we required that a photon's energy and incident angle on the LAT were suitably well reconstructed (event class 3 (diffuse) or better) and any photons
received when \textit{Fermi}'s zenith angle was $<105^{\circ}$ were discarded to
avoid $\gamma-$ray contamination from the Earth.

\subsection{Testing For Deviation from a Simple Power Law}\label{sec:deviation}
Close to the high energy peak in the SED, we expect there to be significant curvature in the spectrum of any given object.  If we have an energy range where a spectrum is well described by a simple power law except for a single break (be it caused by absorption or otherwise) then it is straightforward to identify this break.  However, consider the case where the intrinsic spectrum (i.e. neglecting the break) is so curved that it is not well described by a simple power law.  In this case, the break energy identified by fitting a broken power law would try to describe both the intrinsic curvature and the true break.  In this way the break energy found by the fit is biased from the true break.  Because of this it is important to identify an energy range where curvature in the spectrum is not too great.  We performed Monte Carlo simulations using the standard tool \textit{gtobssim} of an intrinsically curved spectrum with a spectral peak at $\sim250\rm~MeV$ (the spectral parameters were chosen to match the greatest curvature found in the observational sample, $\Gamma=2.11$ and $\beta=0.12$) and introduced an absorption feature above $1.8\rm~GeV$ that a broken power law fit should correctly identify (an optical depth $\tau_{T}=2$ was used which corresponds to $\Delta\Gamma=0.5$ in a spectrum with no curvature).  When we chose the spectral peak as the threshold of the energy range, the break energy was correctly identified ($1.7\pm0.2\rm~GeV$).  However, if we reduced the energy threshold below the spectral peak, the break energy was misidentified (in half the simulations, the break energy varied by more than 1 standard error if the energy threshold was reduced in $50\rm~MeV$ steps).  When we removed the absorption feature from the spectrum, a threshold of $50$~-~$100\rm~MeV$ above the peak was sometimes needed to find a stable break.  The break energy found varied widely between simulations ($3\pm1\rm~GeV$) demonstrating degeneracy in fitting a broken power law to a curved spectrum. These simulations show that if there are spectral breaks in a curved spectrum they can be correctly identified provided the threshold energy is above the spectral peak.  The spectral peaks of each object in our sample have been estimated by \citet{abdo2010c} from the multiwavelength SED. However, there is some uncertainty as with any observation and it is also possible that the spectral peak has shifted with time, meaning the actual peak could be higher than these estimated values.  With this in mind, for each object in our sample, we fitted a broken power law with the threshold first at the peak energy and then $50\rm~MeV$ higher to check the break was stable.  If it was not, we increased the threshold in $50\rm~MeV$ steps until a stable break energy was found. The minimum energy determined in this way was used when fitting all spectral forms. In all cases the threshold was well below where the double-absorber model predicted breaks to occur so they would be identifiable by fitting a broken power law over the appropriate energy ranges (see the following section). Since only bright sources are investigated in this work, the loss of photons incurred by increasing the energy threshold is not detrimental, especially since the containment angle of photons is relatively poor at the low end of the spectrum meaning the signal-to-noise ratio is degraded. In the case of PKS~2022-07 we did not use the peak energy from \citet{abdo2010c} since it could not be estimated directly from the object's SED but could only be estimated from the empirical relationship between spectral index and SED
\begin{equation}
\log (\nu_{\rm{peak}})=-4.0\cdot \Gamma + 3.16.
\end{equation}\newline %
For this object we fitted all of the data above $100\rm~MeV$ with a log-parabola in order to determine the peak energy, which we found to be $290\rm~MeV$.

For each object we took all the photons passing the selection cuts from the low energy threshold up to $100\rm~GeV$.  We then used \textit{gtlike} to determine the likelihood of the data being drawn from the three spectral models: simple power law, broken power law and log-parabola. Because of the low energy threshold imposed, break energies close to this value could not be tested for.  The likelihood calculation incorporated point sources in the region of interest with parameters fixed to those in the \textit{Fermi} 1~Year Catalog \citep{abdo2010b}. Galactic and extragalactic emission were accounted for using the models gll\_iem\_v02 and isotropic\_iem\_v02 respectively\footnote{http://fermi.gsfc.nasa.gov/ssc/data/access/lat/BackgroundModels.html}.  The Akaike Information Criterion~(AIC) of each spectral form was then calculated from its likelihood.  The difference in AIC between models estimates their relative Kullback-Leibler information quantities, i.e. how much each model tested diverges from the distribution of the data relative to the other models \citep{burnham2001}.  The AIC test is particularly suited to multi-model testing as it balances on the one hand the random error in the estimator from overfitting which results from a model containing more free parameters, and on the other hand the systematic error in the estimator from the bias of a model containing fewer parameters \citep{bozdogan1987}. The spectral form with the lowest AIC value is taken to be the best description of the data and differences of $2$ or more between AIC values are considered significant \citep[see][and references therein]{lewis2011}. If a simple power law was not an adequate fit we tested if the break energies were in agreement with those predicted by the double-absorber model and if the break energies were stable with time.

\subsection{Testing For Energy Breaks Caused By BLR Pair-Production}
As noted earlier, it is pointed out in PS10 that a photon-photon absorption feature in a power law spectrum can be approximated as a broken power law.  We confirmed that this approximation is valid by using \textit{gtobssim} to simulate a power law spectrum with photon-photon absorption above a threshold energy and showing that fitting a broken power law using \textit{gtlike} correctly identified this energy.  Because \textit{gtlike} can only fit a power law with a single break, to test the double-absorber model it was necessary to split the photon data from each object into low and high energy sets which were then each fitted with a broken power law.

For the low energy set we calculated the midpoint between the two predicted breaks in the observer frame and took all photons below this energy.  This set was then fitted with a broken power law and tested against a simple power law null-hypothesis. The optimum break energy was then tested to see if it was consistent with the double-absorber model which predicts a break between $4$ and $7\rm~GeV$ in this set. We then selected any photons above the best fit break energy as a high energy set.  We fitted a broken power law to find any optimum second break in the spectrum and tested this against a simple power law null-hypothesis.  Again, we also tested any break to see if it was consistent with the double-absorber model which predicts a second break between $19$ and $30\rm~GeV$.

\subsection{Testing the Stability Of Spectra}
Assuming a break is present in the spectrum of an object, the stability of the break can give important information regarding its cause.  In the double-absorber model the break energy is determined by the wavelength of recombination lines and is therefore predicted to be stable.  Conversely, in the model of \citet{finke2010} the break energy can vary on the timescale of several months as the emission region progresses along the jet.  In general, variation in the spectral shape is expected if it is determined by varying instantaneous physical parameters of the emission region (Klein-Nishina effects, B-field etc.).

We split the dataset from each object into two equal time bins, epoch~1 and epoch~2, and found the spectral shape that best fitted each epoch.  As a test of the stability we compared the break energies of the broken power law fits for each object at the two epochs to see if they were consistent.

\section{Results}
\begin{figure*}
\includegraphics[scale=.20]{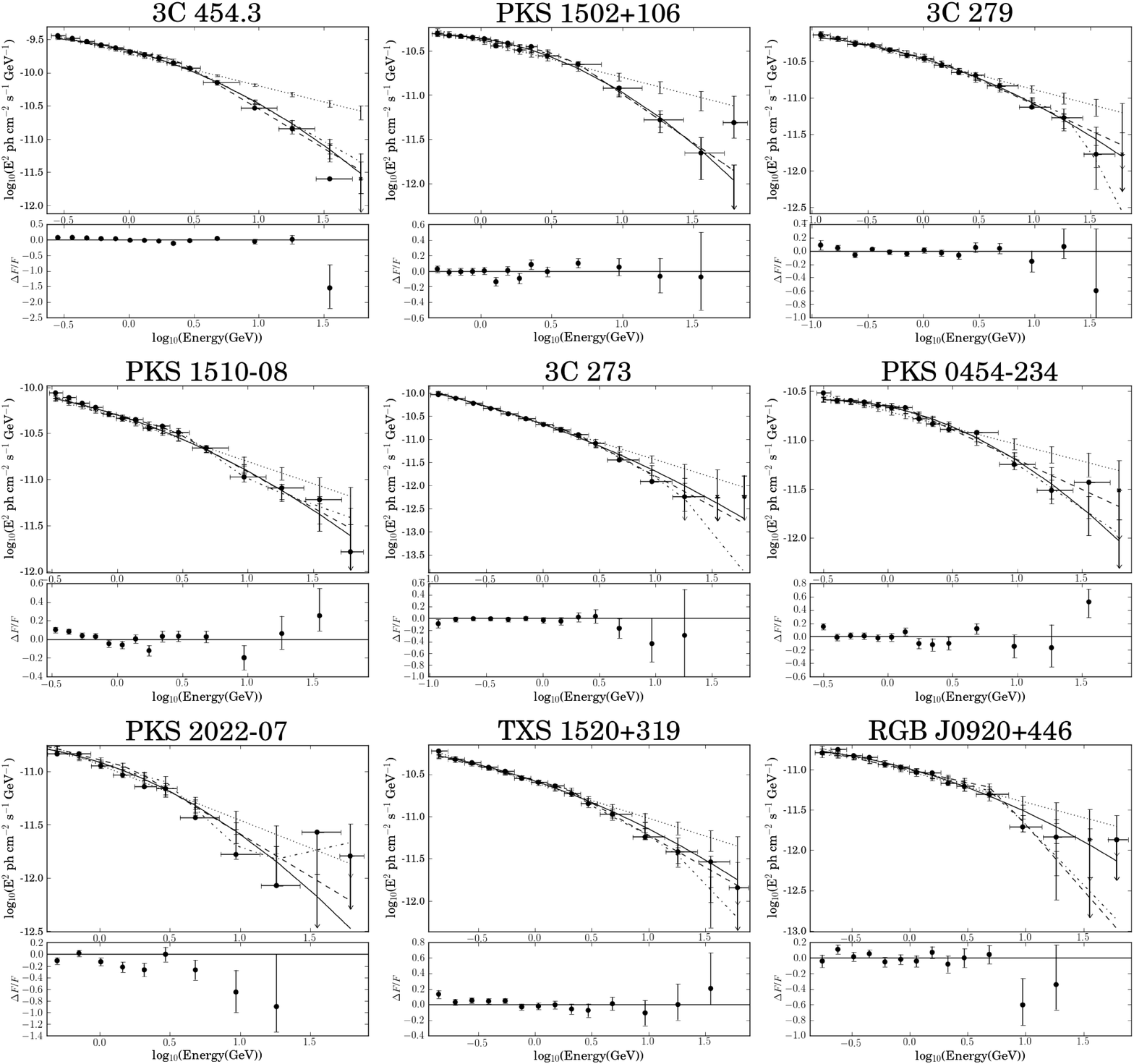}

\caption{\label{fig:ap_phot}Aperture photometry data for each source along with the following spectral shapes from a maximum likelihood fit to the full dataset: simple power law (dotted), broken power law (dashed), log-parabola (solid), and double-absorber (dot-dashed).  Error bars are 68\% confidence intervals for each model (see \citet{aggarwal2012}).  Residuals ((Observed Flux - Model Flux)/Observed Flux) for the best fitting spectral shape are shown underneath each plot.  Energies are in the observer frame.}

\end{figure*}

For illustrative purposes, fits of the different spectral forms to each object are plotted along with aperture photometry data in Figure~\ref{fig:ap_phot}.  The aperture photometry data shows the differential photon flux in energy bins for each object.  For each energy bin we took all photons reconstructed to originate within an on-region of $1^{\circ}$ radius centered on the source.  We then calculated the flux using \textit{gtexposure} to find the average exposure for each energy bin and subtracted an off-region to account for the background.  The off-region was of the same radius as the on-region centered within $15^{\circ}$ from the target and at least $3^{\circ}$ removed from any sources in the \textit{Fermi} 1~Year Catalog \citep{abdo2010b}.

\subsection{Testing For Deviation from a Simple Power Law}
The spectra of all of our sources show significant deviation from a simple power law.  Tables~\ref{simple_table},~\ref{broken_table}, and~\ref{log_table} show the parameters of the best fit to each source using a simple power law, broken power law and log-parabola fit respectively.

\subsection{Testing For Energy Breaks Caused By BLR Pair-Production}
\begin{figure*}
\includegraphics[scale=.20]{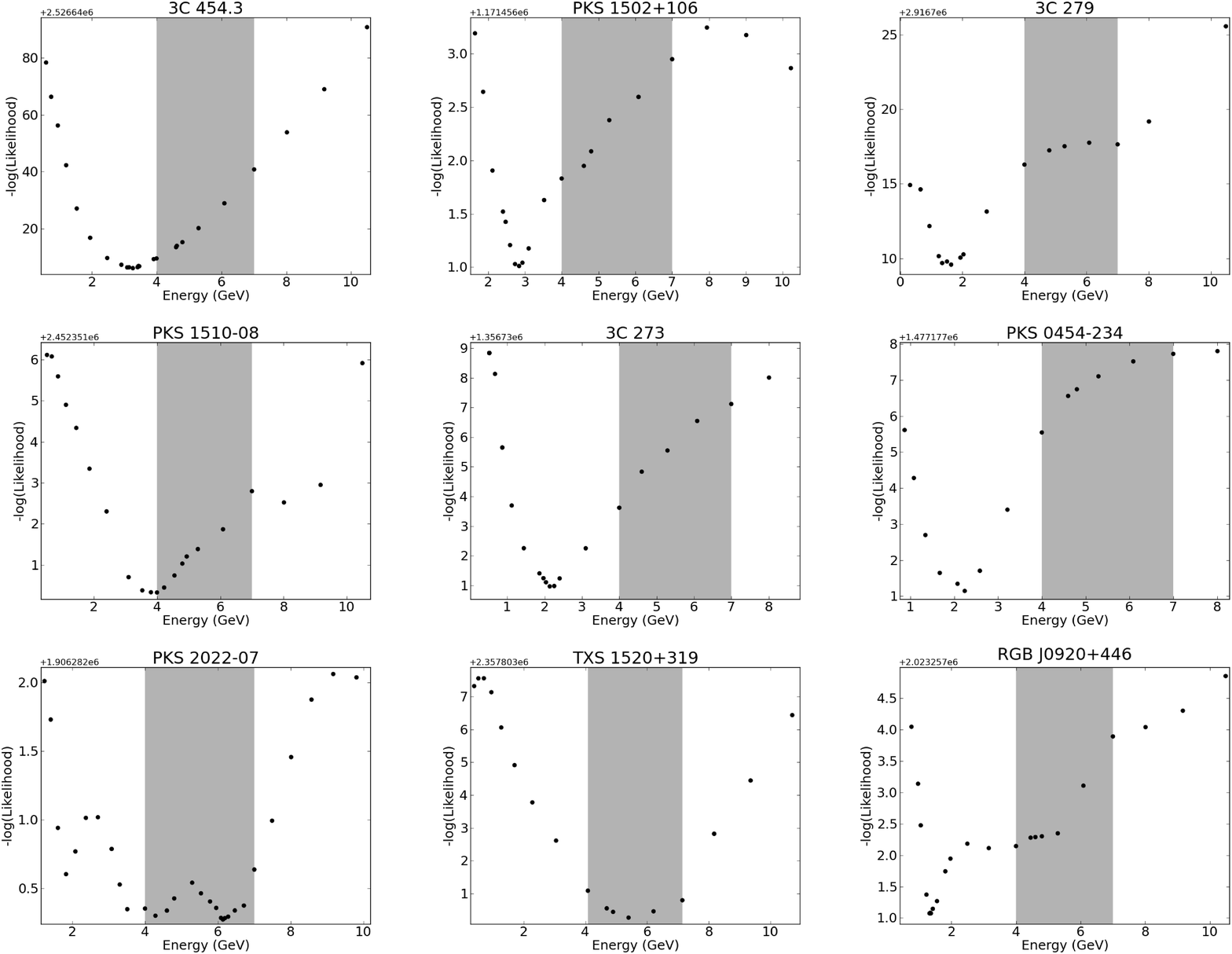}
																			   
\caption{\label{fig:l_montage}Plots of -log(Likelihood) against power law break
energy, in the object's rest frame, for the low energy dataset of each object.  The  shaded box shows the region in which the double-absorber model predicts an energy break to occur.}

\end{figure*}

\begin{figure*}
\includegraphics[scale=.20]{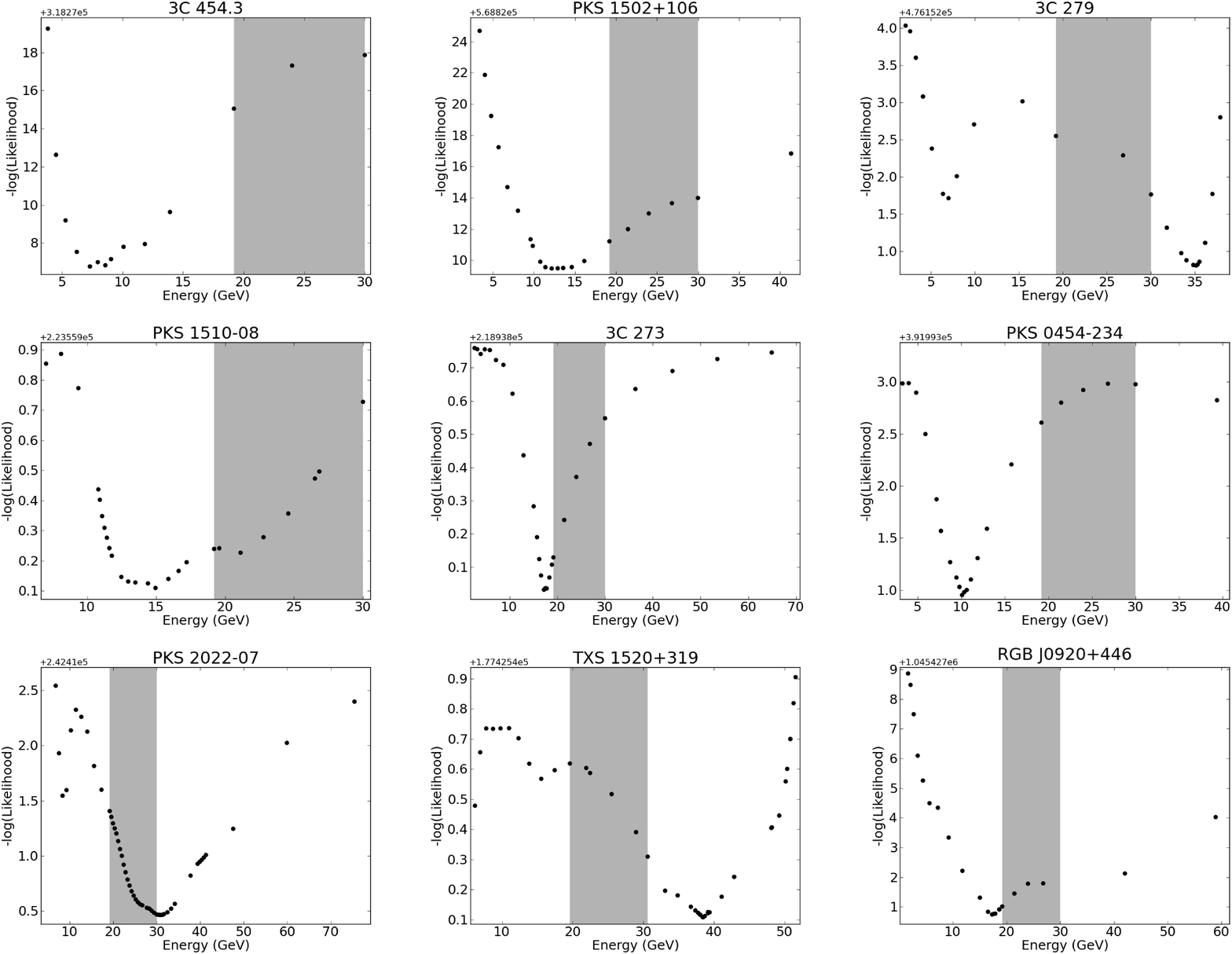}

\caption{\label{fig:h_montage}Plots of -log(Likelihood) against power law break
energy, in the object's rest frame, for the high energy dataset of each object.  The  shaded box shows the region in which the double-absorber model predicts an energy break to occur.}

\end{figure*}

Splitting the dataset for each object into a low energy and high energy set allowed us to test the predictions of the double-absorber model. After splitting the data in this way, the results for broken power law fits to each source and the significance of the improvement they offer over a simple power law are shown in Tables~\ref{LowEnTable} (low energy sets) and~\ref{HighEnTableCon} (high energy sets). Plots of the likelihood of a power law with a given break energy fitting the data are shown in Figures~\ref{fig:l_montage} (low energy sets) and~\ref{fig:h_montage} (high energy sets).  The regions where the double-absorber model predicts breaks to occur are shown as shaded boxes.

When examining the low energy sets, in seven of the nine objects a broken power law is found to be a better description than a simple power law to $>99\%$ significance.  However, in most of the objects, the break energy does not lie within the energy region predicted by the double-absorber model.

Fits to the high energy set of most objects were not significantly improved by using a broken power law.  Of the three objects that do reject the simple power law to $>99\%$ significance, the break energy is again not found in the region predicted by the double-absorber model.  In PS10 evidence of absorption is found for seven objects in which we found a simple power law to be an adequate description, but the results of PS10 were limited by statistics.

\subsection{Testing the Stability Of Spectra}
Results from splitting the data set for each object into two epochs are shown in Table~\ref{StableTable}. Five of the nine sources were best fitted by different spectral shapes in each epoch (i.e. the best fit changed from a log-parabola to a broken power law or vice versa). For the broken power law fits, the break energies in six sources were found to differ over the two epochs by more than one standard deviation, and in three sources they were found to differ by more than two standard deviations. This is more deviation than expected from statistical variation alone, but with a sample size of only nine sources the evidence for variability is weak.

\section{Discussion and Summary}
For each of the nine sources we took all photons with energy below $12\rm~GeV$ in the rest frame of the source and searched for a break.  The double-absorber model predicts that a break should occur in this set between $4$ and $7\rm~GeV$. Our results do not support this because although the fit to the spectra is in most cases significantly improved by introducing a break, the energy at which the break occurs does not fall in the predicted region.  We then took all photons with energies above the low energy break and searched for a second break in the spectrum.  The double-absorber model predicts that such a break should occur between $19$ and $30\rm~GeV$.  Again our results do not support this conclusion because fits to most objects were not significantly improved by introducing a break and in cases where they were the breaks did not fall within the predicted region.

\citet{stern2011} examined the possibility of an intrinsic log-normal spectrum\footnote{They note that this is essentially the same function as a log-parabola in a different representation.} of $\gamma $-rays that contains two `breaks' due to absorption with BLR recombination photons. As described in Section~\ref{sec:deviation}, we ran simulations of a log-parabola with a photon-photon absorption feature and found that fitting a broken power law would correctly identify the break energies in a log-parabola spectrum as proposed by \citet{stern2011}.  Our results do not support a double-absorbed log-parabolic spectrum since the predicted break energies are disfavored in the observational data. For BLR pair-production absorption to remain plausible the photoionization models used in PS10 would require significant modification.  In several of the objects in our sample, a break at $\thicksim1$~-~$3\rm~GeV$ is seen.  To create such a break, a region with a high ionization parameter line-like feature at $\thicksim70$~-~$110\rm~eV$ that dominates over H and He lines would be needed.

When examining all of the data (not splitting the dataset by energy or time), the spectra of all the sources deviate significantly from a simple power law. However, without using physical motivation to prefer one model or the other, a broken power law is not always a better description than a log-parabola.  An AIC test indicates three objects, PKS~1502+106, 3C~279, and RGB~J0920+446, were described significantly better by a log-parabola than a broken power law, and PKS~0454-234 was described marginally better by a log-parabola.  It might be expected that the log-parabola would be favored for weaker sources since with the normalization energy fixed as it is here it has one fewer degree of freedom than a broken power law.  However; these four sources were not the faintest in our sample, in fact 3C~279 had the second greatest flux.  This suggests there is a genuine physical difference in these objects.

For those objects that were better described by a broken power law, the break appears between $\thicksim2$~-~$5\rm~GeV$ in the source rest frame. The break energies determined in the low energy set and with no energy cut are consistent in all objects except 3C~454.3, PKS~1502+106, and RGB~J0920+446, which may suggest the presence of curvature in the spectra of these objects as well as any break. We found weak evidence that the break energies varied with time suggesting that the primary cause of the break is intrinsic to the emission region.

Five of the nine sources are not best fitted by the same spectral shape in both time epochs (i.e. the fit changes from a log-parabola to a broken power law or vice versa).  This could be a change in the spectral shape of the distribution of the particles emitting the spectrum or it could be that when a spectrum with a given spectral shape and time-varying parameters is time-averaged it becomes best fitted by a different spectral shape, a hypothesis that might be testable as more data become available.

Both the wide range of break energies and the variability suggests that the cause of the break cannot be absorption with recombination line photons.  \citet{abdo2010a} note that the break energy is too low to be attributed to Klein-Nishina effects and conclude it is likely caused by a change in the energy distribution of electrons in the jet (assuming a leptonic emission scenario). \citet{finke2010} claim this scenario predicts a correlation between the spectral indices in the optical and $\gamma $-ray regimes, but this is not observed.  They suggest instead two sources of seed photons for external Compton scattering can produce a sharp break, with Klein-Nishina effects providing an important contribution.  In this scenario the break energy changes only gradually as the emission region progresses along the jet; however if a new component was ejected from the central black hole, an event thought to cause flaring in the radio and $\gamma $-ray regimes \citep{acciari2009}, the break energy could be expected to alter rapidly.  This view is consistent with the observations published by \citet{abdo2011}.

In summary, we find evidence that disfavors the double-absorber model for bright \textit{Fermi} FSRQs.  The energy spectrum of $\gamma $-rays from bright FSRQ objects is sometimes better described by a log-parabola and sometimes by a power law with an energy break at a few$\rm~GeV$ which varies with time.  In order to investigate the causes of these features, future work is planned to extend the sample size, to compare FSRQs to BL Lac objects and to investigate any correlations with flux.

\acknowledgments
JH would like to acknowledge provision of a studentship by the UK STFC.  We would like to thank the \textit{Fermi} team for support from their Help Desk and the provision of public data.  We would also like to thank Chris Done, Juri Poutanen and Lowry McComb for helpful discussions throughout writing this paper.

\clearpage

\begin{deluxetable}{lcccc}
\tabletypesize{\scriptsize}
\rotate
\tablecaption{Simple Power Law Fits\label{simple_table}}
\tablewidth{0pt}
\tablehead{
\colhead{Object} &  \colhead{Index} &
\colhead{$\Delta$AIC$_{min}$} \\
\colhead{} & \colhead{$\Gamma$} & \colhead{} \\
\colhead{} & \colhead{} & \colhead{}
}
\startdata
3C 454.3 & $2.485\pm0.006$ & 653.5 \\
PKS 1502+106 & $2.41\pm0.02$ & 78.1 \\
3C 279 & $2.41\pm0.01$ & 70.8  \\
PKS 1510-08 & $2.48\pm0.01$ & 35.6 \\
3C 273 & $2.76\pm0.01$  & 32.9 \\
PKS 0454-234 & $2.34\pm0.02$ & 49.3 \\
PKS 2022-07 & $2.52\pm0.04$ & 7.8 \\
TXS 1520+319 & $2.42\pm0.01$ & 41.6 \\
RGB J0920+446 & $2.35\pm0.02$ & 43.1 \\
\enddata
\tablecomments{The best fitting parameters for simple power law fits to each
object as described in Equation~\ref{simplepl}.
$\Delta$AIC$_{min}$ is the difference in AIC between the simple power law fit
to the object and the best fit out of simple power law, broken power law and
log-parabola, a value of $0$ indicating a simple power law is the best fit to
the data and a value $>2$ indicating either a broken power law or a log-parabola is a significantly better fit to the data.}

\end{deluxetable}

\clearpage

\begin{deluxetable}{lcccccccc}
\tabletypesize{\scriptsize}
\rotate
\tablecaption{Broken Power Law Fits\label{broken_table}}
\tablewidth{0pt}
\tablehead{
\colhead{Object} & \colhead{Break Energy}  &
\colhead{Index1} & \colhead{Index 2} & \colhead{$\Delta$AIC$_{min}$} \\
\colhead{} & \colhead{$E_{b}$} &
\colhead{$\Gamma_{1}$} & \colhead{$\Gamma_{2}$} & \colhead{} \\
\colhead{} & \colhead{(GeV)} & \colhead{} & \colhead{} & \colhead{}
}
\startdata
3C 454.3 & $4.5^{+0.2}_{-0.3}$ & $2.375\pm0.007$ & $3.18\pm0.03$ & 0 \\
PKS 1502+106 & $11.2^{+1.1}_{-0.8}$ & $2.28\pm0.02$ & $3.1\pm0.1$ & 2.6 \\
3C 279 & $2.1^{+0.2}_{-0.3}$ & $2.32\pm0.02$ & $2.70\pm0.04$ & 6.6 \\
PKS 1510-08 & $3.5^{+0.6}_{-0.5}$ & $2.40\pm0.02$ & $2.81\pm0.06$ & 0 \\
3C 273 & $2.2^{+0.2}_{-0.5}$ & $2.72\pm0.02$ & $3.4\pm0.1$ & 0 \\
PKS 0454-234 & $2.6^{+0.5}_{-0.4}$ & $2.12\pm0.04$ & $2.61\pm0.05$ & 1.8 \\
PKS 2022-07 & $4.0^{+1.8}_{-0.7}$ & $2.34\pm0.07$ & $2.8\pm0.1$ & 0 \\
TXS 1520+319 & $4.7^{+1.2}_{-0.6}$ & $2.36\pm0.02$ & $2.76\pm0.06$ & 0 \\
RGB J0920+446 & $16^{+2}_{-5}$ & $2.3\pm0.1$ & $3.6\pm0.4$ & 13.2 \\
\enddata
\tablecomments{The best fitting parameters for broken power law fits to each
object as described in Equation~\ref{brokenpl}.  $E_{b}$ is in the object's rest frame.
$\Delta$AIC$_{min}$ is the difference in AIC between the broken power law fit
to the object and the best fit out of simple power law, broken power law and
log-parabola, a value of $0$ indicating a broken power law is the best fit to
the data and a value $>2$ indicating a log-parabola is a significantly better fit to the data.}
\end{deluxetable}

\clearpage

\begin{deluxetable}{lcccccccc}
\tabletypesize{\scriptsize}
\rotate
\tablecaption{Log-Parabola Fits\label{log_table}}
\tablewidth{0pt}
\tablehead{
\colhead{Object} & \colhead{Index} & \colhead{Curvature} & \colhead{$\Delta$AIC$_{min}$} \\
\colhead{} & \colhead{$\Gamma$} & \colhead{$\beta$} & \colhead{} \\
\colhead{} & \colhead{} & \colhead{} & \colhead{}
}
\startdata
3C 454.3 & $2.307\pm0.009$ & $0.133\pm0.006$ & 12.8 \\
PKS 1502+106 & $2.02\pm0.04$ & $0.16\pm0.02$ & 0 \\
3C 279 & $2.36\pm0.01$ & $0.071\pm0.009$ & 0 \\
PKS 1510-08 & $2.35\pm0.03$ & $0.07\pm0.01$ & 2.4 \\
3C 273 & $2.76\pm0.02$ & $0.07\pm0.01$ & 1.5 \\
PKS 0454-234 & $2.11\pm0.04$ & $0.12\pm0.02$ & 0 \\
PKS 2022-07 & $2.27\pm0.09$ & $0.11\pm0.04$ & 0.4 \\
TXS 1520+319 & $2.38\pm0.01$ & $0.05\pm0.01$ & 9.3 \\
RGB J0920+446 & $2.26\pm0.03$ & $0.10\pm0.02$ & 0 \\
\enddata
\tablecomments{The best fitting parameters for log-parabola fits to each
object as described in Equation~\ref{logparab}.
$\Delta$AIC$_{min}$ is the difference in AIC between the log-parabola fit
to the object and the best fit out of simple power law, broken power law and
log-parabola, a value of $0$ indicating a log-parabola is the best fit to
the data and a value $>2$ indicating a broken power law is a significantly better fit to the data.}
\end{deluxetable}

\clearpage

\begin{deluxetable}{lccccccccc}
\tabletypesize{\scriptsize}
\rotate
\tablecaption{Spectral Properties of Objects over the Whole Time
Period at Low Energy\label{LowEnTable}}
\tablewidth{0pt}
\tablehead{
\colhead{Object} & \colhead{Min Energy} & \colhead{Peak $\rm~log(\nu F_{\nu})$} &
\colhead{$E_{break1}$} & 
\colhead{Index 1} & \colhead{Index 2} & \colhead{Sig over} &
\colhead{$4.8\rm~GeV$} & \colhead{$4$-$7\rm~GeV$} \\
\colhead{} & \colhead{} & \colhead{(GeV)} &
\colhead{(GeV)} & 
\colhead{$\Gamma_1$} & \colhead{$\Gamma_2$} & \colhead{simple PL} & \colhead{exclusion} &
\colhead{exclusion}
}
\startdata
3C 454.3 & 0.46 & 0.244 & $3.3^{+0.2}_{-0.2}$ & 
2.361$\pm$0.008 & 2.86$\pm$0.03 & $>$99\% & $>$99\% & $>99$\% \\
PKS 1502+106 & 1.28 & 0.934 & $2.8^{+0.6}_{-0.4}$ &  2.16$\pm$0.06 &
2.37$\pm$0.05 & 90\% & 86\% & 93\% \\
3C 279 & 0.15 & 0.101 & $1.6^{+0.3}_{-0.4}$ & 
2.30$\pm$0.02 & 2.60$\pm$0.04 & $>$99\% & $>$99\% & $>$99\% \\
PKS 1510-08 & 0.41 & 0.112 & $4.0^{+0.6}_{-1.0}$ &
2.41$\pm$0.02 & 2.9$\pm$0.1 & $>$99\% & 76\% & $<$1\% \\
3C 273 & 0.12 & 0.005  &  $2.1^{+0.2}_{-0.7}$ &
2.70$\pm$0.02 & 3.4$\pm$0.1 & $>$99\% & $>$99\% & $>$99\% \\
PKS 0454-234 & 0.56 & 0.523 & $2.1^{+0.3}_{-0.6}$ &
2.13$\pm$0.04 & 2.45$\pm$0.06 & $>$99\% &  $>$99\% & $>$99\% \\
PKS 2022-07 & 0.24 & 0.693 & $6^{+1}_{-3}$ &
2.38$\pm$0.06 & 3.1$\pm$0.4 & 83\% & 42\% & $<$1\% \\
TXS 1520+319 & 0.29 & 0.253 & $5^{+2}_{-1}$ &
2.36$\pm$0.02 & 3.0$\pm$0.2  & $>$99\% & 45\% & $<$1\% \\
RGB J0920+446 & 0.32 & 0.139 & $1.3^{+0.2}_{-0.2}$ &
2.01$\pm$0.07 & 2.42$\pm$0.04 & $>$99\% & $>$99\% & $>$99\% & \\
\enddata
\tablecomments{Table of results over the whole available time period in the
energy region from Min Energy to $12\rm~GeV$. Values for peak $\rm~log(\nu
F_{\nu})$ taken from~\citet{abdo2010c} except for PKS~2022-07 (see main text). 
$E_{break1}$ is the value for the low energy spectral break which provided the
best fit to the observed data.  $\Gamma_1$ and $\Gamma_2$ are the photon
spectral indices before and after the energy break respectively.  Column 7 shows
how significantly a broken power law improves the fit to the data compared with a
simple power law. Columns 8 and 9 show the confidence level to which one can
exclude energy breaks occurring at $4.8\rm~GeV$ and in the $4-7\rm~GeV$ break
region as predicted in PS10. All energies given in the object's rest frame.}
\end{deluxetable}

\clearpage
\begin{deluxetable}{lccccccccc}
\tabletypesize{\scriptsize}
\rotate
\tablecaption{Spectral Properties of Objects over the Whole Time
Period at High Energy\label{HighEnTableCon}}
\tablewidth{0pt}
\tablehead{
\colhead{Object} & \colhead{Min Energy} &
\colhead{$E_{break2}$ } & 
\colhead{Index 3} & \colhead{Index 4} &\colhead{Sig over} &
\colhead{$19.2\rm~GeV$} & \colhead{$19.2$-$30\rm~GeV$} \\
\colhead{} &  \colhead{(GeV)} & \colhead{(GeV)} &
\colhead{$\Gamma_3$} & \colhead{$\Gamma_4$} & \colhead{simple PL} & \colhead{exclusion} &
\colhead{exclusion}
}
\startdata
3C 454.3 & 3.3 & $7.3^{+1.9}_{-0.7}$ & 2.69$\pm$0.06 & 3.35$\pm$0.7 & $>$99\% & $>$99\% & $>$99\% \\
PKS 1502+106 & 2.8 & $12^{+4}_{-1}$ &2.36$\pm$0.05 & 3.1$\pm$0.1 & $>$99\% & 94\% & $>$99\% \\
3C 279 & 1.6 & $35^{+1}_{-3}$ &2.6$\pm$0.2 & $>$5 & 98\% & 94\% & 95\% \\
PKS 1510-08  & 4.0 & $15^{+13}_{-5}$ &3.0$\pm$0.1 & 2.5$\pm$0.2 & 60\% & 40\% & 53\% \\
3C 273 & 2.1 & $17^{+12}_{-5}$ &3.3$\pm$0.2 & $>$5 & 52\% & 34\% & 47\% \\
PKS 0454-234 & 2.1 & $10^{+2}_{-2}$ &2.50$\pm$0.07& 2.9$\pm0.2$ & 87\% &  93\% & 99\% \\
PKS 2022-07 & 6.0 & $31^{+9}_{-9}$ &3.2$\pm$0.2 & 1.7$\pm$0.5 & 90\% & 83\% & 11\% \\
TXS 1520+319 & 5.0 & $40^{+10}_{-20}$ &2.7$\pm$0.1 & 3.4$\pm$0.6 & 47\% & 69\% & 63\% \\
RGB J0920+446 & 1.3 & $17^{+3}_{-2}$ & 2.39$\pm$0.04 & 3.5$\pm$0.3 & $>$99\% & 53\% & 70\% \\
\enddata
\tablecomments{Table of results over the whole available time period in the
energy region after the first break.  Min Energy is the lowest photon energy for inclusion.
$E_{break2}$ is the value for the high energy spectral break which provided the
best fit to the observed data.  $\Gamma_3$ and $\Gamma_4$ are the photon
spectral indices before and after the energy break respectively. Column 6 shows how significantly a broken power law improves the fit to the data compared with a
simple power law. Columns 7 and 8 show the confidence level to which one can
exclude energy breaks occurring at $19.2\rm~GeV$ and in the $19.2-30\rm~GeV$
break region as predicted in PS10. All energies given in the object's rest frame.}
\end{deluxetable}

\clearpage

\begin{deluxetable}{lcccccccccc}
\tabletypesize{\scriptsize}
\rotate
\tablecaption{Spectral Properties of Objects over different time
periods.\label{StableTable}}
\tablewidth{0pt}
\tablehead{
\colhead{Object} & \colhead{$E_{break}\rm~(GeV)$} & \colhead{$\Delta$AIC$_{min}$} &  \colhead{} & \colhead{$E_{break}\rm~(GeV)$} & \colhead{$\Delta$AIC$_{min}$} &  \colhead{} & \colhead{$E_{break}\rm~(GeV)$} & \colhead{$\Delta$AIC$_{min}$} & \\
\colhead{} & \colhead{No Time cut} & \colhead{} & \colhead{} &
\colhead{Epoch 1} & \colhead{} & \colhead{} & \colhead{Epoch 2} & \colhead{}
}
\startdata
3C 454.3 & $4.5^{+0.2}_{-0.3}$ & 0 & & $4.0^{+0.3}_{-0.3}$ & 0 & & $4.5^{+0.1}_{-0.1}$ & 0.32\\
PKS 1502+106 & $11.2^{+1.1}_{-0.8}$ & 2.6 & & $10.7^{+1.8}_{-0.8}$ & 0.5 & & $2.3^{+0.4}_{-0.3}$ & 0\\
3C 729 & $2.1^{+0.2}_{-0.3}$ & 6.6 & & $1.7^{+0.5}_{-0.1}$ & 6.1 & & $2.4^{+0.6}_{-0.8}$ & 2.9\\
PKS 1510-08 & $3.5^{+0.6}_{-0.5}$ & 0 & & $3.1^{+0.8}_{-0.3}$ & 0 & & $3.9^{+0.5}_{-0.6}$ & 0\\
3C 273 & $2.2^{+0.2}_{-0.5}$ & 0 & & $1.3^{+0.4}_{-0.2}$ & 8.9 & & $4.1^{+0.5}_{-0.5}$ & 0\\
PKS 0454-234 & $2.6^{+0.5}_{-0.4}$ & 1.8 & & $3.3^{+0.3}_{-0.5}$ & 1.5 & & $1.4^{+0.2}_{-0.2}$ & 0\\
PKS 2022-07 & $4.0^{+1.8}_{-0.7}$ & 0 & & $3.6^{+1.1}_{-0.9}$ & 0 & & $5.6^{+2.7}_{-0.9}$ & 0\\
TXS 1520+319 & $4.7^{+1.2}_{-0.6}$ & 0 & & $4.4^{+2.2}_{-0.6}$ & 0.7 & & $4.4^{+2.0}_{-0.5}$ & 0\\
RGB J0920+446 & $16^{+2}_{-5}$ & 0 & & $1.0^{+0.4}_{-0.1}$ & 0 & & $5.3^{+0.6}_{-1.2}$ & 0\\
\enddata
\tablecomments{Table of results found by splitting the data for each object
into two equal time epochs.  The most likely break energy for each object, in the object's rest frame, with no time cut and in each epoch individually is shown along with $68\%$ confidence intervals and the difference from the AIC minimum, a value of $0$ indicating a broken power law is the best fit to the data and a value $>2$ indicating a log-parabola is a significantly better fit to the data.}
\end{deluxetable}

\clearpage

\clearpage

\clearpage

\clearpage

\end{document}